\documentclass[twocolumn]{revtex4}

\usepackage{epsfig}
\usepackage{amssymb}
\usepackage{amsmath}
\usepackage{amsfonts}
\usepackage{graphicx}
\usepackage{mathrsfs}
\usepackage[dvips]{color}
\usepackage{multirow}
\usepackage{graphicx}
\usepackage{epstopdf}

\newcommand{\R}{\mathbb{R}}
\newcommand{\C}{\mathbb{C}}

\newcommand{\fz}{\mathfrak{z}}

\newcommand{\fK}{\mathfrak{K}}

\newcommand{\bbe}{\mathbf{e}}

\newcommand{\bk}{\mathbf{k}}

\newcommand{\bbr}{\mathbf{r}}

\newcommand{\bH}{\mathbf{H}}
\newcommand{\bI}{\mathbf{I}}

\newcommand{\bM}{\mathbf{M}}

\newcommand{\cM}{\mathcal{M}}

\newcommand{\cP}{\mathcal{P}}

\newcommand{\cT}{\mathcal{T}}

\newcommand{\cX}{\mathcal{X}}

\newcommand{\be}{\begin{equation}}
\newcommand{\ee}{\end{equation}}
\newcommand{\bea}{\begin{eqnarray}}
\newcommand{\eea}{\end{eqnarray}}
\newcommand{\nn}{\nonumber}

\newcommand{\ed}{\end{document}}

\newcommand{\bi}{\begin{itemize}}
\newcommand{\ei}{\end{itemize}}

\newcommand{\bce}{\begin{center}}
\newcommand{\ece}{\end{center}}

\newcommand{\sE}{\mathscr{E}}
\newcommand{\sF}{\mathscr{F}}

\newcommand{\sT}{\mathscr{T}}

\begin{document}

\title{Exactness of the Born Approximation and Broadband Unidirectional\\ Invisibility in Two Dimensions}


\author{Farhang Loran$^*$ and Ali~Mostafazadeh$^\dagger$\\[6pt]
$^*$Department of Physics, Isfahan University of Technology, Isfahan, 84156-83111, Iran\\[6pt]
$^\dagger$Departments of Mathematics and Physics, Ko\c{c} University, 34450 Sar{\i}yer,
Istanbul, Turkey}

\begin{abstract}
Achieving exact unidirectional invisibility in a finite frequency
band has been an outstanding problem for many years. We offer a
simple solution to this problem in two dimensions that is based on
our solution to another more basic open problem of scattering
theory, namely finding potentials ${v(x,y)}$ whose scattering
problem is exactly solvable via the first Born approximation.
Specifically, we find a simple condition under which the first Born
approximation gives the exact expression for the scattering
amplitude whenever the wavenumber for the incident wave is not
greater than a given critical value ${\alpha}$. Because this
condition only restricts the ${y}$-dependence of ${v(x,y)}$, we can
use it to determine classes of such potentials that have certain
desirable scattering features. This leads to a partial inverse
scattering scheme that we employ to achieve perfect
(non-approximate) broadband unidirectional invisibility in two
dimensions. We discuss an optical realization of the latter by
identifying a class of two-dimensional isotropic active media that
do not scatter incident TE waves with wavenumber in the range
${(\alpha/\sqrt 2,\alpha]}$ and source located at ${x=\infty}$,
while scattering the same waves if their source is relocated to
${x=-\infty}$.
\end{abstract}

\maketitle


\section{Introduction}

In 1926, the year Born published his monumental work on the
probabilistic interpretation of quantum mechanics \cite{born1}, he
also laid the foundations of quantum scattering theory and
introduced the celebrated Born approximation \cite{born2}. The
latter proved to be an extremely powerful tool for performing
scattering calculations in different areas of physics
\cite{hofstadter,koshino,born-wolf}. The Born approximation of order
$N$ corresponds to the approximation scheme in which one neglects
all but the first $N+1$ terms in the standard series solution of the
Lippmann-Schwinger equation \cite{sakurai}. For a scattering
potential that is proportional to a coupling constant $\fz$, this
leads to an approximate expression for the scattering amplitude $f$
which is a polynomial  $f_N$ of degree $N$ in $\fz$; $N$-th order
Born approximation corresponds to $f\approx f_N$.

Because of its central importance, Born approximation is discussed
in standard textbooks on quantum mechanics \cite{sakurai,griffiths},
optics \cite{born-wolf}, and scattering theory \cite{newton}. But,
surprisingly, none of these address the natural problem of inquiring
into potentials for which the $N$-th order Born approximation gives
the exact expression for the scattering amplitude, i.e., $f=f_N$. In
this article, we offer a solution of this problem for the case $N=1$
in two dimensions, i.e., identify potentials in two dimensions for
which the first Born approximation is exact. This may seem as a
purely academic problem, but its solution proves to have
far-reaching consequences; it paves the way for devising a method of
engineering scattering potentials which we employ to address another
outstanding problem of basic importance, namely, achieving perfect
unidirectional invisibility in a tunable finite frequency band.

The study of invisible potentials has been a subject of research for
many decades. In one dimension, a potential is invisible from the
left (respectively right), if it does not reflect a left-
(respectively right-) incident plane wave and transmits it without
changing the amplitude or phase of the wave, i.e., it is
transparent. Reciprocity theorem implies that the transparency
cannot be unidirectional \cite{bookchapter}. A nonreal scattering
potential can, however, support unidirectional reflectionlessness
and invisibility \cite{lin,invisible3-1,invisible3-2,invisible3-3}.
The principal example is
    \be
    v(x)=\left\{\begin{array}{ccc}
    \fz\,e^{2i\beta x} &{\rm for} & x\in[0,L],\\
    0  &{\rm for} & x\notin[0,L],\end{array}\right.
    \label{exp-pot}
    \ee
where $\fz$ is a coupling constant, $\beta$ is a nonzero real
parameter, and $L:=\pi/|\beta|$,
\cite{lin,invisible-1,invisible-2,invisible-3}.  This potential is
unidirectionally invisible from the right (respectively left) for an
incident wave with wavenumber $k=|\beta|$, if  $\beta>0$
(respectively $\beta<0$) and $\beta^2|\fz|\ll 1$,
\cite{lin,longhi-2011,jones}. The latter condition is to ensure the
validity of the first Born approximation. For sufficiently large
values of $\beta^2|\fz|$, the first Born approximation is unreliable
and the unidirectional invisibility of the potential (\ref{exp-pot})
breaks down \cite{pra-2014a}.

The problem of realizing exact unidirectional invisibility is
addressed in
Refs.~\cite{pra-2014a,ge-2012,pra-2013a,ap-2014,pra-2014b,jpa-2016}.
Similarly to (\ref{exp-pot}), the unidirectionally invisible
potentials considered in these references display this property for
a single or a discrete set of wavenumbers. In one dimension, the
problem of constructing potentials that have exact unidirectional
invisibility in the entire wavenumber spectrum is solved in
\cite{horsley-2015,longhi-2015}. See also
\cite{longhi-op-2015,longhi-op-2016,jiang,hayran}.

Ref.~\cite{prsa-2016} generalizes the notion of unidirectional
invisibility at isolated values of the wavenumber to two and three
dimensions and examines its approximate realization for weak
potential where
the first Born approximation is relible. A more recent development
is the construction of a class of scattering potentials in two
dimensions that display exact broadband omnidirectional invisibility
\cite{ol-2017}; these do not scatter incident plane waves with an
arbitrary incidence angle provided that their wavenumber $k$ does
not exceed a prescribed value.  The broadband invisibility achieved
in \cite{ol-2017} is bidirectional in the sense that the scattering
amplitude vanishes regardless of the position of the source of the
incident wave. The quest for realizing exact unidirectional
invisibility in an extended frequency band, which we address in this
article, has been a well-known open problem for many years. This is
a much more important problem, because its solution would allow for
the design of material with broadband nonreciprocal functionalities.

\section{Exactness of Born approximation}

Consider a two-dimensional scattering setup where the scatterer is
described by a possibly complex-valued potential $v(x,y)$. The
source of the incident wave, which is considered to be a plane wave,
can be placed at $x=-\infty$ or $x=+\infty$. We use the terms
``left-incident'' and ``right-incident waves'' to refer to these
cases, respectively. In general, the wave vector $\bk_0$ of the
incident wave makes an angle $\theta_0$ with the $x$-axis, i.e.,
$\bk_0=k (\cos\theta_0 \bbe_x+\sin\theta_0\bbe_y)$, where $\bbe_j$
is the unit vector along the $j$-axis with $j=x,y$. We use
$\theta_0^{l/r}$ to label $\theta_0$ for the left/right-incident
waves. Clearly,
    \begin{align}
    &0<\cos\theta^l_0\leq 1, && -1\leq\cos\theta^r_0<0.
    \label{cos-cos}
    \end{align}

By definition, if $v(x,y)$ is a scattering potential, the solutions
of the Schr\"odinger equation,
    \be
    [-\nabla^2+v(x,y)]\psi(x,y)=k^2\psi(x,y),
    \label{sch-eq}
    \ee
tend to plane waves at spatial infinities. In particular,
(\ref{sch-eq}) admits the so-called ``scattering solutions,"
$\psi^{l/r}$, that satisfy
$\psi^{l/r}(\bbr)=e^{i\bk_0\cdot\bbr}+\sqrt{i/kr}\,e^{ikr}
f^{l/r}(\theta)$ {\rm as} $r\to\infty$, where
$\bbr:=x\,\bbe_x+y\,\bbe_y$, $(r,\theta)$ are the polar coordinates
of $\bbr$, and $f^{l/r}(\theta)$ is the scattering amplitude for the
left/right-incident waves \cite{adhikari-86}. The latter stores the
scattering properties of the potential. Therefore its determination
is the main objective of the scattering theory.

It is not difficult to show that the first Born approximation yields
\cite{prsa-2016}:
    \be
    f^{l/r}(\theta)=-\frac{
    \tilde{\tilde v}\big(\mbox{\small $k(\cos\theta-\cos\theta^{l/r}_0),
    k(\sin\theta-\sin\theta^{l/r}_0)$}\big)}{2\sqrt{2\pi}},
    \label{f=2}
    \ee
where $\tilde{\tilde v}(\fK_x,\fK_y):=\int_{-\infty}^\infty dx
\int_{-\infty}^\infty dy\, e^{-i(x\fK_x+y\fK_y)}v(x,y)$ is the
two-dimensional Fourier transform of $v(x,y)$. We wish to find
conditions under which (\ref{f=2}) gives the exact expression for
the scattering amplitudes of the potential. Our main technical tool
for achieving this purpose is the transfer-matrix formulation of
potential scattering in two dimensions \cite{pra-2016}. We summarize
its basic ingredients in the sequel.\\[6pt]
\noindent 1) For a given wavenumber $k$, let $\sF_k$ denote the
space of complex-valued functions $\xi$ such that $\xi(p)=0$ for
$|p|\geq k$;
    \[\sF_k:=\{\,\xi:\R\to\C\: |\:\xi(p)=0~{\rm for}~|p|\geq k\,\}.\]
The transfer matrix of a scattering potential $v(x,y)$ is a $2\times
2$ matrix $\bM$ with operator entries $M_{ij}$ acting in $\sF_k$.\\[6pt]
\noindent 2) We can express $\bM$  as the time-ordered exponential
of a non-Hermitian effective $2\times 2$ matrix Hamiltonian $\bH(x)$
with operator entries $H_{ij}(x)$ acting in $\sF_k$;
    \begin{align}
    \bM=&\sT\exp\left\{-i\int_{-\infty}^\infty dx\;\bH(x)\right\}
    :=\bI-i\int_{-\infty}^\infty dx\;\bH(x)+\nn\\
    &
    (-i)^2\int_{-\infty}^\infty dx_2\int_{-\infty}^{x_2} dx_1\;\bH(x_2)\bH(x_1)+\cdots,
    \label{time-ordered}
    \end{align}
where $\sT$ is the time-ordering operation with $x$ playing the role
of time, and $\bI$ is the identity operator for the space $\sF_k^2$
of two-component state vectors with components belonging to $\sF_k$.
The entries of $\bH(x)$ are defined by
    \be
    H_{ij}(x)\xi(p):=\frac{\epsilon_i e^{-i\epsilon_i\varpi(p)x}}{2\varpi(p)}
    \,v(x,i\partial_p)\!\left[e^{i\epsilon_j\varpi(p)x}\xi(p)\right],
    \label{H-comp}
    \ee
where $\epsilon_i:=(-1)^{i-1}$, $\varpi(p):=\sqrt{k^2-p^2}$,
$v(x,i\partial_p)$ is the linear operator acting in $\sF_k$
according
    to
            \be
            v(x,i\partial_p)\xi(p):=\frac{1}{2\pi}\int_{-k}^k dq\,
            \tilde v(x,p-q) \xi(q),
            \label{v-dp}
            \ee
    and $\tilde v(x,\fK_y):=\int_{-\infty}^\infty dy\,e^{-i\fK_y y}v(x,y)$
    is the Fourier transform of $v(x,y)$ with respect to $y$.\\[3pt]\\[6pt]
\noindent 3) The scattering amplitudes $f^{l/r}$ are given by
        \be
        f^{l/r}(\theta)=\frac{-ik|\cos\theta|}{\sqrt{2\pi}}\times \left\{
                \begin{array}{cc}
                T^{l/r}_-(k\sin\theta) &{\rm for}~ \cos\theta<0,\\
                T^{l/r}_+(k\sin\theta) &{\rm for}~  \cos\theta>0,
                \end{array}\right.
                \label{f=}
                \ee%
    where $T^{l/r}_\pm$ are the elements of $\sF_k$ fulfilling
        \begin{align}
        &M_{22}T_-^{l}(p)=-2\pi M_{21}\delta(p-p_0^l),
        \label{Tm-L}\\
        &T^l_+(p)=
        M_{12}T_-^l(p)+2\pi(M_{11}-I)\delta(p-p_0^l),
        \label{Tp-L}\\
        &M_{22}T_-^{r}(p)=-2\pi (M_{22}-I)\delta(p-p_0^r),
        \label{Tm-R}\\
        &T^{\rm r}_+(p)
        =M_{12}[T^r_-(p)+2\pi\delta(p-p_0^r)],
                \label{Tp-R}
        \end{align}
    $\delta(\cdot)$ stands for the Dirac delta function,
    $p_0^{l/r}:=k\sin\theta_0^{l/r}$,
    and $I$ is the identity operator acting in $\sF_k$.\vspace{6pt}

Because  $M_{22}$ is in general an integral operator, (\ref{Tm-L})
and (\ref{Tm-R}) are linear integral equations. According to
(\ref{f=}), we can solve the scattering problem for the potential
$v(x,y)$ provided that we determine the transfer matrix $\bM$ and
solve (\ref{Tm-L}) and (\ref{Tm-R}).
Refs.~\cite{pra-2016,pra-2017,jpa-2018} offer details of the
application of this scheme for solving concrete scattering problems.

The transfer matrix $\bM$, the functions $T^{l/r}_\pm$, and the
scattering amplitude $f^{l/r}$ depend on
the wavenumber $k$. A critical observation underlying the present
study is that under a fairly simple condition on the potential, the
Dyson series for the transfer matrix (\ref{time-ordered}) truncates
for wavenumbers not exceeding a critical value $\alpha$;
    \be
    \bM=\bI-i\int_{-\infty}^\infty dx\,\bH(x)~~{\rm for}~~k\leq\alpha.
    \label{thm1}
    \ee
Furthermore, the same condition allows for an explicit solution of
the integral equations yielding $T^{l/r}_-$. As we show in
Appendix~A, this provides explicit formulas for $T^{l/r}_\pm$ and
$f^{l/r}(\theta)$ for $k\leq\alpha$, and proves:\vspace{3pt}

\noindent {\em Theorem~1}. Let $v(x,y)$ be a scattering potential
satisfying
    \be
    \tilde v(x,\fK_y)=0~~{\rm for}~~\fK_y\leq \alpha,
    \label{condi-1}
    \ee
where $\alpha$ is a given positive real parameter. Then the first
Born approximation provides the exact solution of the scattering
problem for $v(x,y)$ whenever the wavenumber $k$ of the incident
wave does not exceed $\alpha$.\vspace{3pt}

\noindent This result is reminiscent of the notion of a
quasi-exactly solvable potential \cite{turbiner}. The
time-independent Schr\"odinger equation for such a potential can be
solved to determine finitely many low-lying bound state energies and
the corresponding eigenfunctions. The potentials fulfilling
(\ref{condi-1}) may also be viewed as quasi-exactly solvable,
because their scattering problem is exactly solvable for energies
$k^2\leq \alpha^2$. Notice also that according to a result of
Ref.~\cite{ol-2017}, condition (\ref{condi-1}) implies
omnidirectional invisibility of the potential for $k\leq\alpha/2$
for both left- and right-incident waves, i.e., for these values of
$k$, $f^{l/r}(\theta)=0$. Theorem~1 is a much stronger result,
because it provides an explicit formula for the scattering
amplitudes $f^{l/r}(\theta)$ also for $k\in(\alpha/2,\alpha]$ where
they need not vanish.

Let us also note that Theorem~1 does not imply that the first Born
approximation gives the exact solution of the Schr\"odinger equation
(\ref{sch-eq}) for $k\leq\alpha$. It is indeed not difficult to show
that the second and higher order terms in the Born series expansion
of the wave function are generally nonzero, but for $k\leq\alpha$
they yield evanescent waves which do not contribute to the
scattering amplitude of the potential.

\section{Perfect broadband unidirectional invisibility}
Consider the scattering of left- and right-incident waves 
by a potential $v(x,y)$ satisfying (\ref{condi-1}), and suppose that
$k\leq\alpha$. Then (\ref{f=2}) holds, and (\ref{cos-cos}) implies
    \bea
    &&-2\alpha \leq -2k\leq k(\cos\theta-\cos\theta_0^l)\leq k\leq\alpha,
    \label{L-inv-1}\\
    &&-\alpha<-k\leq k(\cos\theta-\cos\theta_0^r)<2k\leq 2\alpha.
    \label{L-inv-2}
    \eea
According to (\ref{f=2}) and (\ref{L-inv-1}), $f^l(\theta)=0$ for
all possible values of $\theta^l_0$ and $\theta$ provided that
$\tilde{\tilde v}(\fK_x,\fK_y)=0$ for $\fK_x\in[-2\alpha,\alpha)$.
In other words this is a sufficient condition for the
left-invisibility of $v(x,y)$ whenever $k\leq\alpha$. For scattering
potentials of physical interest, this condition is equivalent to
    \be
    \tilde v(\fK_x,y)=0~~{\rm for}~~ \fK_x\in[-2\alpha,\alpha),
    \label{condi3-1}
    \ee
where $\tilde v(\fK_x,y):=\int_{-\infty}^\infty
dx\:e^{-i\fK_xx}v(x,y)$ is the Fourier transform of $v(x,y)$ with
respect to $x$. Similarly, we can use (\ref{L-inv-2}) to obtain the
following sufficient condition for the right-invisibility of the
potential for $k\leq\alpha$.
    \be
    \tilde v(\fK_x,y)=0~~{\rm for}~~ \fK_x\in(-\alpha,2\alpha].
    \label{condi3-4}
    \ee
Because (\ref{condi3-1}) and (\ref{condi3-4}) do not coincide, there
is a range of values of $k$ for which only one of these conditions
holds. This provides the basic motivation for achieving broadband
unidirectional invisibility. It leads to the following theorem whose
proof we give in Appendix~B.
\vspace{3pt}\\
\noindent {\em Theorem~2.} A scattering potential $v(x,y)$ is
unidirectionally right- (respectively left-) invisible for
wavenumbers $k\in(\alpha/\sqrt 2,\alpha]$, if it satisfies
(\ref{condi-1}), (\ref{condi3-4}), and
    \be
    \tilde v(\fK_x,y)\neq 0~~{\rm for}~~\fK_x\in[-2\alpha,-\alpha],
    \label{condi3-2}
    \ee
(respectively (\ref{condi-1}), (\ref{condi3-1}), and $\tilde
v(\fK_x,y)\neq 0$ for $\fK_x\in[\alpha,2\alpha]$.)\vspace{3pt}

The conditions appearing in the hypothesis of Theorems~1 and 2 are
not difficult to satisfy. In Appendix~C we construct an infinite
class of scattering potentials fulfilling these conditions. A simple
example is
    \be
    v(x,y):=\frac{\fz\, e^{i\alpha(y-x)}}{
    \left[\left(x/a- i\right)\left(y/b+i\right)\right]^{2}},
    \label{pot=1}
    \ee
where $\fz$ is a real or complex coupling constant, and $a$ and $b$
are real parameters. For all values of $\fz$, $a$, and $b$, this
potential displays unidirectional right-invisibility whenever the
wavenumber of the incident wave is in the range $(\alpha/\sqrt
2,\alpha]$. To see this, we evaluate the Fourier transform of the
right-hand side of (\ref{pot=1}) with respect to $y$ and use the
result to check that it fulfills (\ref{condi-1}). According to
Theorem~1, this shows that we can employ (\ref{f=2}) to calculate
the scattering amplitudes of (\ref{pot=1}) for $k\leq\alpha$. The
result of this calculation is
    \be
    f^{l/r}(\theta)=-\pi\sqrt{2\pi}\: \fz\, a^2b^2k^2
    \cX(ak,c^{l/r})\: \cX(bk, s^{l/r}),
    \label{f-left=1}
    \ee
where
    \begin{align*}
    &\cX(\xi,\zeta):=\left\{
    \begin{array}{ccc}
    \zeta e^{-\xi\zeta} & {\rm for} & \zeta> 0\\
    0 & {\rm for} &\zeta\leq 0\end{array}\right.,\\
    &s^{l/r}:=\sin\theta-\sin\theta_0^{l/r}-\alpha/k,\\
    &c^{l/r}:=\cos\theta_0^{l/r}-\cos\theta-\alpha/k.
    \end{align*}
Because, $\cos\theta_0^r<0$, for $k\leq\alpha$ we have $c^r<0$. This
implies that $\cX(ak, c^{r})=0$. Therefore, $f^r(\theta)=0$ for all
$\theta\in[0,2\pi)$, i.e., the potential (\ref{pot=1}) is
right-invisible.  It is easy to check that for
$\theta_0^l\in(-\pi/2,0)$ there are ranges of values of $\theta$
within the interval $(\pi/2,\pi)$ for which $f^l(\theta)\neq 0$.
Therefore, the right-invisibility of the potential is
unidirectional.

\section{Optical realization}

Complex scattering potentials in two dimensions may be used to model
the scattering of transverse electric (TE) waves by the
inhomogeneities of an effectively two-dimensional nonmagnetic
isotropic medium $\cM$, \cite{born-wolf,ol-2017}. Let
$\varepsilon(x,y)$ label the permittivity profile of $\cM$ and
suppose that as $r\to\infty$, $\varepsilon(x,y)$ tends to a constant
value $\varepsilon_\infty$. For a TE wave propagating in $\cM$, we
can express its electric field in the form $\sE_0e^{-i\omega
t}\psi(x,y)\bbe_z$, where $\sE_0$ is a constant amplitude, $t$
labels time, $\bbe_z$ is the unit vector along the $z$-axis,
$\omega:=\sqrt{\varepsilon_0/\varepsilon_\infty}\,ck$,
$\varepsilon_0$ and $c$ are respectively the permittivity and the
speed of light in vacuum, and $k$ is the wavenumber. We can use
Maxwell's equations to identify $\psi(x,y)$ with a solution of the
Schr\"odinger equation (\ref{sch-eq}) for the potential
    \be
    v(x,y)=k^2[1-\hat\varepsilon(x,y)],
    \label{v-opt}
    \ee
where $\hat\varepsilon(x,y):=\varepsilon(x,y)/\varepsilon_\infty$ is
the relative permittivity of $\cM$.

In order for the medium to display perfect broadband unidirectional
invisibility for wavenumbers $k\in(\alpha/\sqrt 2,\alpha]$, it
suffices to select a reasonable value for $\varepsilon_\infty$,
identify the left-hand side of (\ref{v-opt}) with one of the
potentials fulfilling the hypothesis of Theorem~2, and solve this
equation for $\hat\varepsilon(x,y)$.

\begin{figure}
    \begin{center}
    \includegraphics[scale=.60]{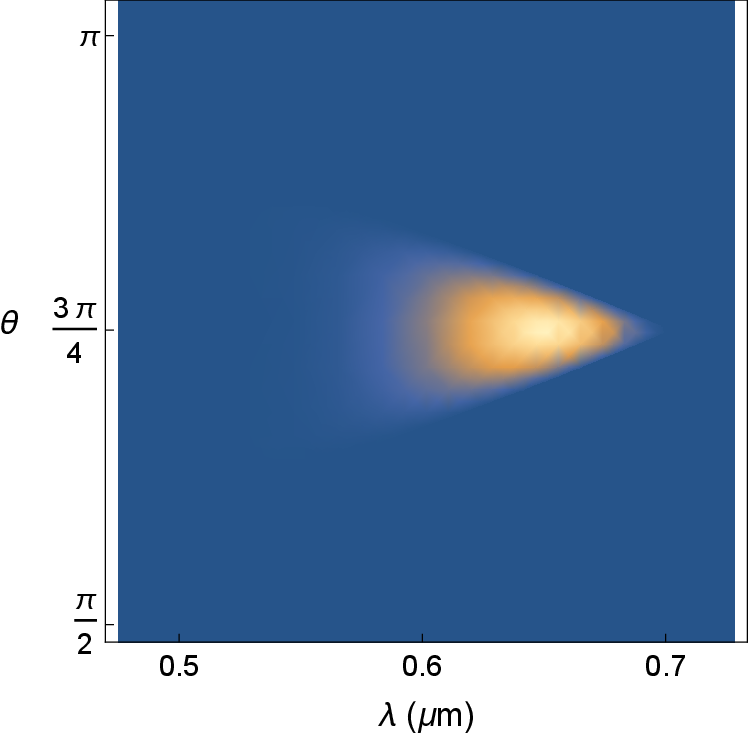}
    \caption{Density plot of $|f^l|^2$ as a function of the polar angle $\theta$ and the wavelength $\lambda$ of a left-incident wave with incident angle $\theta^l_0=-\pi/4$ for permittivity profile (\ref{eps=1}) with
$\alpha=2\pi/500~{\rm nm}$ and $a=b=1~\mu{\rm m}$. Regions colored
by dark blue are those where $f^l(\theta)=0$. Lighter colors
represent regions with larger values of  $|f^l(\theta)|^2$. These
lie to the left of the line
$\lambda=2\sqrt{2}\pi/\alpha=0.707~\mu{\rm m}$.}
    \label{fig1}
    \end{center}
    \end{figure}

For example consider the relative permittivity profile corresponding
to the potential (\ref{pot=1}), i.e.,
    \be
    \hat\varepsilon(x,y)=1+\frac{\fz_{-}e^{i\alpha(y-x)}}{
    \left[\left(x/a- i\right)
    \left(y/b+i\right)\right]^{2}},
    \label{eps=1}
    \ee
where $\fz_-:=-\fz/k^2$ is a free dimensionless coupling constant.
By construction, Eq.~(\ref{f-left=1}) gives an exact description of
the scattering of TE waves by the optical medium possessing the
relative permittivity profile (\ref{eps=1}) whenever their
wavenumber does not exceed $\alpha$. As we noted above, this
equation establishes the right-invisibility of the medium for TE
waves with wavenumber $k\leq\alpha$ (and their superpositions).
Fig.~\ref{fig1} provides a graphical demonstration of the
unidirectionality of the right-invisibility of this medium for
wavenumbers in the range $(\alpha/\sqrt 2,\alpha]$. Here we have set
$\theta_0^l=-\pi/4$, $\alpha=2\pi/500~{\rm nm}$, and $a=b=1~\mu{\rm
m}$.
As expected, the medium described by (\ref {eps=1}) has a nonzero scattering amplitude for left-incident waves with incidence angle $\theta_0^l=-\pi/4$ provided that their wavelength $\lambda:=2\pi/k$ lies between 500.0~nm and 707.1~nm. \\[6pt]

\section{Concluding remarks}
In this article we have offered a solution of the old problem of
finding potentials whose scattering problem can be solved exactly
via the first Born approximation. These provide the first examples
of quasi-exactly solvable scattering potentials, because the first
Born approximation gives the exact expression for their scattering
amplitudes whenever the wavenumber of the incident wave  does not
exceed a prescribed value $\alpha$.

Condition (\ref{condi-1}) that ensures the exactness of the first
Born approximation for wavenumbers $k\leq\alpha$ does not restrict
the $x$-dependence of the potential $v(x,y)$. Because, according to
(\ref{f=2}), such a potential may be obtained by performing the
inverse Fourier transform of the scattering amplitudes
$f^{l/r}(\theta)$, we can fix the $x$-dependence of the potential by
demanding that $f^{l/r}(\theta)$ has a certain desirable behavior
for $k\leq\alpha$. This yields an extremely effective partial
inverse scattering prescription which we have presently employed for
achieving perfect broadband unidirectional invisibility. The latter
admits optical realizations involving certain two-dimensional
nonmagnetic isotropic media with regions of gain and loss. We expect
a two-dimensional analog of the setup employed in Ref.~\cite{jiang}
to allow for an experimental detection of this effect.

The generalization of our results to three dimensions does not pose
any major difficulty. In particular, one can pursue the
above-mentioned idea of partial inverse scattering in three
dimensions. The progress in this direction can have interesting
applications in acoustics.

\section*{Appendix~A: Proof of Theorem~1}

The proof of Theorem~1 rests on (\ref{thm1}) and the fact that we
can use this equation to obtain closed-form expressions for
$T_\pm^{l/r}(p)$ and $f^{l/r}(\theta)$ whenever $k\leq\alpha$ . In
the following we outline a derivation of these expressions and
provide a proof of Theorem~1.

We begin by noting that for each $p\in[-k,k]$, $-k+p\leq 0<\alpha$.
Using this inequality in  (\ref{v-dp}), we find
    \begin{align}
    &v(x,i\partial_p)\xi(p)=\frac{1}{2\pi}\int_{\alpha}^{k+p}\! dq\: \tilde v(x,q)\xi(p-q).
    \label{vx}
    \end{align}
To establish (\ref{thm1}), we set $k\leq\alpha$ and use (\ref{vx})
to show that for all $\xi_1,\xi_2\in\sF_k$,
    \[v(x_2,i\partial_p)[\xi_2(p)v(x_1,i\partial_p)\xi_1(p)]=0.\]
This together with (\ref{H-comp}) imply
    \be
    H_{ij}(x_2)H_{{i'}{j'}}(x_1)\xi(p)=0~~{\rm for}~~k\leq\alpha.
    \label{condi-2}
    \ee
Hence $\bH(x_2)\bH(x_1)$  vanishes, and (\ref{time-ordered}) implies
(\ref{thm1}).

Next, we note that according to (\ref{thm1}) and (\ref{condi-2}),
    \be
    (M_{ij}-\delta_{ij}I)(M_{i'j'}-\delta_{i'j'}I)=0~~{\rm for}~~k\leq\alpha,
    \label{thm2}
    \ee
where $\delta_{ij}$ denotes the Kronecker delta symbol, and  $0$
stands for the zero operator acting in $\sF_k$. A simple consequence
of (\ref{thm2}) is a straightforward calculation of $T^{l/r}_\pm$.
To see this, we express (\ref{Tm-L})  and (\ref{Tm-R}) in the form
    \bea
    T_-^l(p)&=&-(M_{22}-1)T_-^l(p)-2\pi M_{21}\delta(p-p_0^l),
    \label{Tm-L-2}\\
    T_-^r(p)&=&-(M_{22}-1)[T_-^r(p)+2\pi \delta(p-p_0^r)].
    \label{Tm-R-2}
    \eea
Similarly to (\ref{Tp-L})  and (\ref{Tp-R}), the right-hand sides of
these equations consist of terms obtained by applying operators of
the form $M_{i'j'}-\delta_{i'j'}I$ to functions belonging to
$\sF_k$. In light of (\ref{thm2}), this implies
$(M_{ij}-\delta_{ij}I)T^{l/r}_\pm(p)=0$. Using this relation in
(\ref{Tp-L}), (\ref{Tp-R}), (\ref{Tm-L-2}), and (\ref{Tm-R-2}), we
obtain
    \bea
    T_-^l(p)&=&-2\pi M_{21}\delta(p-p_0^l),
    \label{Tm-L-3}\\
    T^l_+(p)&=&2\pi (M_{11}-I)\delta(p-p_0^l),
    \label{Tp-L-3}\\
    T_-^r(p)&=&-2\pi (M_{22}-1)\delta(p-p_0^r),
    \label{Tm-R-3}  \\
    T^r_+(p)&=&2\pi M_{12}\delta(p-p_0^r).
    \label{Tp-R-3}
    \eea
To obtain more explicit formulas for $T^{l/r}_\pm(p)$, we introduce
    \[h_{ij}(p,q):=\tilde{\tilde v}\big(\epsilon_i\varpi(p)-\epsilon_j\varpi(p-q),q\big),\]
and use (\ref{H-comp}), (\ref{vx}), and (\ref{thm1}) to show that
    \[(M_{ij}-\delta_{ij})\xi(p)=\frac{-i \epsilon_i}{4\pi\varpi(p)}\int_\alpha^{k+p}
    \!\!dq\, h_{ij}(p,q)\,\xi(p-q).\]
Substituting this equation in (\ref{Tm-L-3}) -- (\ref{Tp-R-3}) and
using the result together with the identities, $p=k\sin\theta$ and
$\varpi(p)=k|\cos\theta|$, in (\ref{f=}), we obtain (\ref{f=2}).
This proves Theorem~1.

\section*{Appendix~B: Proof of Theorem~2}

Suppose that (\ref{condi-1}) and (\ref{condi3-4}) hold. Then
(\ref{f=2}) applies, and the potential is right-invisible for all
$\theta_0^r$ and $k\leq\alpha$. To ensure that it is not
left-invisible, we must determine values of $k$ within the interval
$(0,\alpha]$ and ranges of values of $\theta_0^l$ and $\theta$ for
which $f^l(\theta)\neq 0$, i.e., $\tilde{\tilde
v}\big(k(\cos\theta-\cos\theta^l_0),k(\sin\theta-\sin\theta^l_0)\big)\neq
0$. Conditions (\ref{condi-1}) and (\ref{condi3-4}) violate this
inequality unless $\sin\theta-\sin\theta_0^l>\alpha/k$ and
$\cos\theta_0^l-\cos\theta\geq\alpha/k$. We have used these
relations together with (\ref{cos-cos}) to show that
    \begin{align}
    &-\frac{\pi}{2}<\theta_0^l<0, &&\frac{\pi}{2}\leq\theta<\pi,
    &&\frac{\alpha}{\sqrt 2}<k\leq\alpha,
    \nn
    \end{align}
and $\varphi_k<\theta-\theta^l_0<2\pi-\varphi_k$, where
$\varphi_k:=2\arcsin\left( \alpha/\sqrt
2\,k\right)\in\big[\pi/2,\pi\big)$. Similarly, we find the following
necessary conditions for unidirectional left-invisibility of
$v(x,y)$: $\varphi_k<\theta^r_0-\theta<2\pi-\varphi_k$ and
    \begin{align}
    &\pi<\theta_0^r\leq \frac{3\pi}{2}, &&0<\theta\leq\frac{\pi}{2},
    &&\frac{\alpha}{\sqrt 2}<k\leq\alpha.
    \nn
    \end{align}
This completes the proof of Theorem~2.

\section*{Appendix~C: Construction of potentials displaying broadband unidirectional
 invisibility}

 It is easy to see that under multiplication of a function $g(x)$ by $e^{i\beta x}$
its Fourier transform $\tilde g(\fK)$ changes to $\tilde
g(\fK-\beta)$. This allows for expressing conditions appearing in
the statement of Theorems~1 and~2 in terms of functions with
vanishing Fourier transform with respect to $x$ or $y$ on the
negative or positive  $x$- or $y$-axes \cite{footnote1}. To be more
specific, let $w_\pm(x,y)$ and $v_{l/r}(x,y)$ be scattering
potentials satisfying
    \begin{align}
    &\tilde w_\pm(\fK_x,y)=0~~{\rm respectively~for}~~\pm\fK_x\leq 0,
    \label{condi-w1}\\
    &\tilde w_\pm(x,\fK_y)=0~~{\rm for}~~\fK_y\leq 0,
    \label{condi-w3}\\
    &v_l(x,y)=e^{i\alpha y}\left[
        \gamma e^{-i2\alpha x}w_-(x,y)+e^{i\alpha x}w_+(x,y)\right],
    \label{v-L}\\
    &v_r(x,y)=e^{i\alpha y}\left[
        e^{-i\alpha x}w_-(x,y)+\gamma e^{2i\alpha x}w_+(x,y)\right],
    \label{v-R}
    \end{align}
    where $\gamma=0,1$. Then, according to Theorem~2, for $k\in(\alpha/\sqrt 2,\alpha]$, $v_l$ is unidirectionally left-invisible, if
        \bea
        \tilde w_+(\fK_x,y)\neq 0&{\rm for}& \fK_x\in[0,\alpha],
        \label{condi-w4}
        \eea
    and $v_r$ is unidirectionally right-invisible, if
        \bea
        \tilde w_-(\fK_x,y)\neq 0&{\rm for}& \fK_x\in[-\alpha,0].
        \label{condi-w5}
        \eea

We can construct specific examples for $w_\pm(x,y)$ using functions
of the form
    \[g_j(x):=\left(x/L_j+i\right)^{-n_j-1},\]
   where $j=1,2,3,4$, and $L_j$ and $n_j$ are respectively positive real numbers and positive integers. It is easy to check that $\tilde g_j(\fK)=0$ for $\fK\leq 0$. This in turn implies that the function $\bar g_j(x):= g_j(x)^*$ satisfies $\tilde{\bar g}_j(\fK)=0$ for $\fK\geq 0$. In view of these properties of $g_j$ and $\bar g_j$, we can show that the following choices for $w_\pm(x,y)$ fulfill (\ref{condi-w1}), (\ref{condi-w3}), (\ref{condi-w4}), and (\ref{condi-w5}).
    \begin{align}
    &w_-(x,y)=\fz_-\bar g_1(x)g_2(y),
    &&w_+(x,y)=\fz_+g_3(x)g_4(y),
    \label{wpm=}
    \end{align}
where $\fz_\pm$ are nonzero real or complex coupling constants.
Substituting (\ref{wpm=}) in (\ref{v-L}) and (\ref{v-R}), we
therefore find scattering potentials that are unidirectionally left-
and right-invisible for $k\in(\alpha/\sqrt 2,\alpha]$. It is
important to note that this is true for arbitrary choices of $L_j$
and $n_j\geq 1$, and that linear combinations of potentials of this
form (with different choices for $L_j$ and $n_j$) posses the same
unidirectional invisibility property.

Now, consider the $w_-(x,y)$ of (\ref{wpm=}) with $g_1(x)$ and
$g_2(y)$ given by  $n_1=n_2=1$, and let $a:=L_1$ and $b:=L_2$.
Substituting this choice for $w_-(x,y)$ in (\ref{v-R}) and setting
$\gamma=0$ yield the potential (\ref{pot=1}) with $\fz:=-k^2\fz_-$.

\subsection*{Acknowledgements}
We are indebted to the Turkish Academy of Sciences (T\"UBA) for
providing the financial support for FL's visit to Ko\c{c} University
during which we have carried out a major part of the work reported
here. AM has been supported by T\"UBA's principal membership grant.


\ed